\documentstyle[12pt, aaspp4,epsfig]{article}
\def\DLya{{DLA\ }}
\begin{document}

\title{\hspace{0.5in} The $z=0.0912$ and $z=0.2212$ Damped Ly$\alpha$
Galaxies Along the Sight-Line Toward the Quasar OI$\;$363\footnotemark}
\footnotetext{Based on observations obtained with the Mayall
4.0-m NOAO Telescope on Kitt Peak, operated for NSF by AURA, the 3.5-m WIYN
Telescope on Kitt Peak, also operated for NSF by AURA (WIYN is a joint
facility of University of Wisconsin, Indiana University, Yale University, 
and NOAO), the NASA 3.0-m IRTF on Mauna Kea, operated for NASA by University 
of Hawaii, and the Hiltner 2.4-m Telescope on Kitt Peak, operated by MDM Observatory
(this is a joint facility of University of Michigan, Dartmouth College,
Ohio State University, and Columbia University).}

\author{David A. Turnshek,\altaffilmark{2}
Sandhya Rao,\altaffilmark{2}
Daniel Nestor,\altaffilmark{2}
Wendy Lane,\altaffilmark{2,3} 
Eric Monier\altaffilmark{2,4}}

\medskip

\affil{Department of Physics \& Astronomy, University of Pittsburgh,
Pittsburgh, PA 15260, USA}

\author{Jacqueline Bergeron,\altaffilmark{2}}

\medskip

\affil{ESO, Karl-Schwarzschild-Strasse 2, Garching bei Munchen, D-85748,
Germany}

\and

\author{Alain Smette\altaffilmark{2}}

\medskip

\affil{NASA Goddard Space Flight Center, Code 681, Greenbelt, MD 20771,
USA }

\altaffiltext{2}{email: turnshek@quasar.phyast.pitt.edu,
rao@everest.phyast.pitt.edu, dbn@phyast.pitt.edu, 
wlane@astro.rug.nl, monier@astronomy.ohio-state.edu, jbergero@eso.org,
asmette@band3.gsfc.nasa.gov}

\altaffiltext{3}{also Kapteyn Astronomical Institute, University of
Groningen (present address).}

\altaffiltext{4}{also Ohio State University (present address).}

\begin{abstract}

New optical and infrared observations along the sight-line toward the
quasar OI$\;$363 (0738+313) are presented and discussed.  Excluding
systems which lack confirming UV spectroscopic observations of the
actual Ly$\alpha$ line, this sight-line presently contains the two
lowest-redshift classical damped Ly$\alpha$ (\DLya) quasar absorption line
systems known (i.e. with $N_{HI} \ge 2 \times 10^{20}$ atoms cm$^{-2}$),
one at $z_{abs}=0.0912$ and the other at $z_{abs}=0.2212$. Our new
observations suggest identifications for the \DLya galaxy counterparts of
these absorption-line systems. The $z=0.09$ \DLya galaxy appears to be an
extended low surface brightness galaxy which is easily visible only in
infrared images and shows rich morphological structure.  Assuming there
is no contribution from the quasar host galaxy, we place an upper limit on
the K-band luminosity of the $z=0.09$ \DLya galaxy of $L_K \le 0.13L^*_K$
(assuming a cosmology with $H_0=65$ km s$^{-1}$ Mpc$^{-1}$, $\Omega=1$,
and $\Lambda=0$).  More realistically, a subtraction of the quasar nuclear
and host light yields $L_K \approx 0.08L^*_K$.  The impact
parameter between the galaxy and quasar sight-line is very small, $b<3.6$
kpc ($<2$ arcsec), which makes measurements difficult.  The $z=0.22$
\DLya galaxy is an early-type dwarf with a K-band luminosity of L$_K
\approx 0.1L^*_K$ at impact parameter $b=20$ kpc.  Its colors are
neutral and consistent with star formation models suggesting its formation
epoch was less than a few Gyr ago (i.e. $z_f\approx0.3-0.9$).  Thus,
it is conceivable that its progenitor originated from the population of
``faint blue galaxies'' seen at moderate redshifts. In general, these
results serve to support mounting evidence that \DLya galaxies are drawn
from a wide variety of gas-rich galaxy types.

\end{abstract}


\newpage

\section{INTRODUCTION}

The $z_{abs}=0.0912$ and $z_{abs}=0.2212$ damped Ly$\alpha$ (DLA)
absorption-line systems along the sight-line toward the quasar OI$\;$363
(0738+313) have been the subject of several recent papers by members
of our group. Rao \& Turnshek (1998, hereafter RT98) announced the
discovery of these systems during the course of an HST-FOS survey for
\DLya absorption in QSO MgII systems.\footnotemark \footnotetext{The
results of the full spectroscopic survey for low-redshift \DLya systems
are discussed in Rao \& Turnshek (2000).} In this paper they discussed the
results of a preliminary optical imaging search for galactic light from
these two absorbers, and they concluded that the galactic counterparts
are definitely not luminous spirals.  Lane et al. (1998a,b) and Lane,
Briggs \& Smette (2000a) presented the results of 21 cm observations
along the sight-line toward OI$\;$363. They found that both of the damped
systems exhibit 21 cm absorption. However,  an initial search for 21
cm emission from the lower-redshift system failed to detect a signal
(Lane et al. 2000b).

The existence of these two systems is noteworthy because they currently 
represent the two lowest-redshift classical quasar \DLya absorption-line
systems which are confirmed, i.e., actual UV spectroscopic observations of
their \DLya profiles exist and analyses of the spectra demonstrate that
the systems' neutral hydrogen column densities are above the classical
survey limit of $N_{HI} \ge 2 \times 10^{20}$ atoms cm$^{-2}$. In several
other reported cases of low-redshift \DLya systems, either confirming UV
spectroscopy is lacking (Miller, Knezek \& Bregman 1999), the \ion H1 column
density is somewhat below the classical limit (Lanzetta et al. 1997),
or both. Thus, although detection of foreground galaxies near a bright
background QSO remains difficult, the OI$\;$363 sight-line offers a
most ideal opportunity for studying any emission from  stellar and gaseous 
components of objects responsible for classical \DLya absorbers.
We refer to such luminous objects as \DLya galaxies.

Here we report on the results of new observations of galaxies along the
sight-line toward OI$\;$363.  The data consist of optical and infrared
imaging and also optical spectroscopy in seeing conditions sometimes
as good as $\approx 0.55$ arcsec. Throughout the analysis we assume $H_o =
65$ km s$^{-1}$ Mpc$^{-1}$, $\Omega=1$, and $\Lambda=0$; we also adopt the
naming convention used in RT98. In \S2 we start by summarizing the results
from our earlier studies.  In \S3 the new observations and analyses
of these data are presented.  In brief, the \DLya galaxy associated
with the $z=0.0912$ absorber appears to be an extended low surface
brightness dwarf galaxy at impact parameter $b<3$ kpc, with rich
morphological structure and a K-band luminosity of $L_K\le0.08-0.13L^*_K$,
where the measured light is an upper limit depending on the amount of
contamination from the quasar host galaxy
at redshift $z \approx 0.63$ ($L_K\le0.13L^*_K$ assumes no host contribution).
The \DLya galaxy associated with the
$z=0.2212$ absorber is considerably different in character; it is an
early-type dwarf galaxy which is relatively compact at an impact parameter
of $b=20$ kpc, with a K-band luminosity of $L_K\approx0.1L^*_K$.
In \S4 we discuss the implications of these findings.

\section{PREVIOUS RESULTS}

RT98 presented the HST-FOS UV spectrum of OI$\;$363 (see their figures
1-3) which led to the identification of the two absorbing systems at
$z_{abs}=0.0912$ and $z_{abs}=0.2212$ with \ion H1 column densities of $N_{HI}
= (1.5\pm0.2) \times 10^{21}$ atoms cm$^{-2}$ and $N_{HI} = (7.9\pm1.4)
\times 10^{20}$ atoms cm$^{-2}$, respectively.

RT98 also presented a 45 min R-band WIYN image of the OI$\;$363 field
in seeing of 0.55 arcsec (see their figure  4). Analysis of this
image resulted in a list of 15 resolved objects and 11 point sources
brighter than R = 24 mag within 40 arcsec of the quasar sight-line (see
their table 1). The closest resolved object was found to be 
a galaxy (named G1,
R = 20.8 mag) 5.7 arcsec to the south-southeast of the quasar and the
closest unresolved object is presumably a star (named S1, R = 20.8 mag)
2.5 arcsec to the northeast of the quasar. Fainter objects were visible
in the R-band WIYN image but they were not tabulated since such a list
would have suffered from incompleteness.  However, the existence of
a marginally significant fainter extended feature a few arcsec to the
west-southwest of the quasar sight-line was noted (marked with a cross on
figure 4 of RT98). It appears to be the brightest part of the ``fuzz''
near the quasar, and was reported to have an R-band surface brightness
of $\approx 25.3$ mag arcsec$^{-2}$. Notably, the brightest galaxy in
the field is a luminous spiral (named G11, R = 17.1 mag) $\approx 31$
arcsec to the east of the quasar with a redshift of $z \approx 0.06$. A
similarly luminous spiral at a somewhat larger redshift of $z=0.09$, i.e.,
the redshift of the lowest redshift \DLya absorber, could have easily
been detected in this field if it was present, but the observations ruled
this out.  We note that there is an absorption line at $\approx1290$ \AA\
in the original FOS low-resolution spectrum of the quasar, which can be
identified as Ly$\alpha$ with a rest equivalent width of W$\approx0.7$
\AA\ and a redshift of $z_{abs}\approx0.06$. The existence of a Ly$\alpha$
line at $z=0.06$ is consistent with the results of Guillemin \& Bergeron 
(1996).  In any case,
due to their close proximity to the quasar sight-line in comparison to
other objects in the field, the galaxy labeled G1 and any object that
might be associated with the fuzz are of most significant interest;
the results from our new observations (\S3) indeed suggest that these
luminous objects can be identified with the two \DLya galaxies.

Both of the \DLya absorbers along this sight-line have been shown to
have associated \ion H1 21cm absorption (Lane et al. 1998a,b and 2000a;
Chengalur \& Kanekar 1999).  The $z_{abs}=0.2212$ system shows only a
single narrow 21 cm absorption feature, while the $z_{abs}=0.0912$ system
has complex velocity structure.  The spectrum from this lower redshift
system is best fit by multiple Gaussians: two narrow cold components
with thermal kinetic widths of T$_k \approx 300$ and $100$ K, and a
third broad component with T$_k \approx  5000$ K (Lane et al. 2000a).
This suggests the presence of two temperature phases for the neutral
medium, similar to those found in the Galaxy. For this lower redshift
system, excluding the velocity interval where absorption is detected
(a region covering $\approx 55$ km s$^{-1}$), an upper limit has been
placed on the \ion H1 gas mass.  Sensitive WSRT 21 cm emission
measurements of the $z_{abs}=0.0912$ absorber place the upper limit at
$M_{HI} \leq 3.7 \times 10^9$h$_{65}^{-2}$ M$_{\odot}$ for
an assumed velocity spread of 100 km s$^{-1}$ (Lane et al. 2000b). This
requires the \ion H1 mass of the absorber to be somewhat less than that of
a normal spiral galaxy.  In principal, since the velocity spread would
be lower for a more face-on spiral, the formal \ion H1 mass limit for this
special case could be even lower.  However, if the 21 cm emission were
confined to a very narrow velocity interval, one might then be concerned
that a significant amount of the 21 cm emission were missed because it
fell in the same velocity intervals as the absorption.

\section{OBSERVATIONS AND ANALYSES OF NEW DATA}

New observations along the sight-line toward OI$\;$363 include optical and
infrared imaging and optical spectroscopy. Based on our identifications,
here we exclusively consider analysis of the data pertaining to
galaxy G1 and the fuzz near the quasar.  G1 is shown to be a dwarf
galaxy at redshift $z\approx0.22$, consistent with the redshift of the
$z_{abs}=0.2212$ \DLya absorber.  We argue that at least part of the
fuzz is associated with the $z_{abs}=0.0912$ \DLya absorber, leading
to the interpretation that the $z=0.09$ \DLya galaxy is an extended low
surface brightness dwarf galaxy with rich morphological structure.

\subsection{Optical Imaging}

Optical BRI imaging data of the OI$\;$363 field were obtained for us by
the WIYN Queue team using a $2048\times2048$ Tektronics STIS thinned,
frontside illuminated CCD (0.195 arcsec per pixel) on the 3.5-m WIYN
Telescope on Kitt Peak.  In addition, we obtained supplemental U-band
imaging data using the $1024\times1024$ Templeton thinned, backside
illuminated CCD (0.28 arcsec per pixel) on the MDM Observatory 2.4-m
Hiltner Telescope on Kitt Peak. The seeing ranged between $\approx
0.55-1.0$ arcsec. The journal of observations for these optical images
is presented in Table 1.  These data were processed using normal
methods. Observations of standard stars and field stars in the quasar
frame allowed us to obtain calibrated magnitudes on the Johnson (U, B)
and Cousins (R, I) system.  We note that the R calibration is consistent
with the calibration used in RT98 which was originally based on the
calibration of Drinkwater, Webster, \& Thomas (1993).  However, the four
color images are not shown here because they do not reveal any additional
qualitative information beyond what can be learned from inspection of
the R-band WIYN image that was published by RT98.

\subsection{Infrared Imaging}

Infrared imaging observations of the OI$\;$363 field in JHK were obtained
on the 3.0-m NASA IRTF on Mauna Kea using NSFCAM, a 256$\times$256 InSb
detector array (0.30 arcsec per pixel).  The journal of observations for
these infrared images is also presented in Table 1.  The observations
were obtained during the course of three different observing runs in
various observing conditions. The conditions were not always photometric 
and there were variations in the infrared sky brightness; the seeing 
ranged between $0.75-1.0$ arcsec.  Using recommended procedures (see the
irtf.ifa.hawaii.edu/Facility/nsfcam/nsfcam.html WEB site), the
observations were obtained using several different short-exposure dither
patterns, flatfielded with sky frames that were obtained from the
dithered object frames, and then shifted before addition
to obtain a ``final'' image.  Since several different standard stars
were used and the data were collected on a variety
of dates, this provided multiple opportunities to check the photometric
calibrations.  Panels (a) and (b) of
Figure 1 show one of the K band images (60 min) of the OI$\;$363 field
and its associated isophotal plot. This was one of the best K-band 
images in the sense that it had good seeing ($\approx$ 0.75 arcsec) 
and a low K-band sky brightness, allowing very low surface brightness
features to be revealed. For display, this image was smoothed with a Gaussian 
(FWHM = 0.75 arcsec) and the unresolved nuclear PSF was then subtracted
(see \S3.4.2 for discussion of this process). 
Due to poorer seeing and/or higher background 
sky levels, the remaining K-band images provided little additional 
information, with the exception of the checks of the 
photometric calibration.

\subsection{Optical Spectroscopy of G1}

Optical spectroscopic observations of objects in the OI$\;$363 field
were obtained using the NOAO 4.0-m Mayall Telescope on Kitt Peak. The aim
of the spectroscopy was to identify or exclude objects near the \DLya
redshifts of $z=0.22$ and $z=0.09$. We will only present the
spectrum of G1 here since the fuzz near the quasar was too faint to obtain
a reliable spectrum and the other spectra we obtained did not result in 
identifications for either of these two redshifts.

Spectra of objects in this field were obtained with the CRYOCAM in
long-slit and multi-slit modes. Since G1 is both faint and close to the
bright QSO, obtaining its spectrum was difficult.  We found that the
long slits produced stronger object signals that were easier to measure
relative to sky than the short multi-slit masks.  The final spectrum of
G1 is a combination of two long-slit spectra obtained separately with
the Grism 650 (400 l mm$^{-1}$ covering $4000-6800$ \AA) and the Grism
770 (300 l mm$^{-1}$ covering $4300-8500$ \AA). The bluer spectrum was
obtained in 120 minutes on the night of 29 April 1998 with a 1.7 arcsec
wide slit at a resolution of $\approx 10$ \AA\ (3.2 \AA\ per pixel);
the redder spectrum was obtained in 90 minutes on the night of 14 Feb
1999 with a 1.0 arcsec wide slit at a resolution of $\approx 12$ \AA\
(4.3 \AA\ per pixel). The slit width used was matched to the seeing
conditions as closely as possible.  Wavelength calibration of the spectra
was accomplished with the observation of an internal HeNeAr lamp. The
G1 spectra were extracted from the flatfielded Loral (Ford) CCD images
after removal of cosmic rays using an optimal extraction algorithm.  The
spectra were initially flux-calibrated using standard star observations,
adjusted to the same flux level in the overlapping region and combined
by variance weighting.  The observations produced a useful spectrum
of G1 over the wavelength interval $4500-8500$ \AA. The $f_{\lambda}$
spectrum, rebinned to $\approx$ 12 \AA\ pixels and normalized in flux
at 6000 \AA, is shown in Figure 2. The downturn in the flux beyond 8000
\AA\ is possibly due to poor calibration at the red end of the Grism
770 spectrum where the spectrograph becomes more defocussed.

\subsection{The Properties of Two Low-Redshift \DLya Galaxies}

The IRTF K-band image (Figure 1) shows G1 ($\approx5.7$ arcsec southeast of
the quasar) and reveals the presence of luminous features in the fuzz 
surrounding the quasar that were not very significant in the RT98 WIYN 
R-band image.  The fuzz is seen to consist of features of low, but 
not uniform, surface brightness.  One gets the impression that, within the
fuzz, the elongated feature to the east-southeast of the quasar could be
part of a (spiral) ``arm'' since it is approximately concentric with respect
to the somewhat off-centered light surrounding the quasar; 
a similar arm is not apparent 
on the opposite side of the quasar. To the west-southwest
there is a feature elongated from east to west that appears like a ``jet''
since it runs nearly perpendicular to the otherwise concentric-looking
extended light. This jet-like feature appears to have structure.
It is coincident with the cross marked on the original RT98
WIYN R-band image, but it does not emanate from the quasar nucleus. In
the discussion below we will use the terms ``arm'' and ``jet'' to describe
the appearance of these features, but there is certainly no {\it strong}
evidence that these terms describe their actual physical nature.

Table 2 lists the UBRIJHK photometric measurements of G1, the arm, and
the jet.  Details of how these measurements were made and how they have
been used to infer the properties of the two \DLya galaxies at $z=0.22$
and $z=0.09$ are given below. Since we have a spectrum of G1, we will
consider it first.

\subsubsection{G1 ($z=0.22$)}

\medskip

\centerline{\it 3.4.1.1. The Spectrum}

\smallskip

The spectrum of G1 is shown in Figure 2.  For comparison, the spectrum of
an early-type galaxy (the type E3 elliptical galaxy NGC 4648, Kennicutt
1992a), redshifted to $z=0.22$ and normalized at 6000 \AA, is also
plotted. This template spectrum is offset relative to the spectrum of
G1 for clarity. The spectrum of this early-type galaxy template over our
observed wavelength range is dominated by the spectral features of cool
giants, i.e., the 4000 \AA\ break, the G band at rest wavelength 4304 \AA,
the MgI $+$ MgH (Mgb) band at 5175 \AA, the NaD lines at 5893 \AA, and
a TiO band at 6260 \AA.  The positions of these features as they appear
in the redshifted spectrum of NGC 4648 are marked.  The telluric B band
in this spectrum also appears redshifted, since it was not removed from
the original Kennicutt (1992a) spectrum. The unredshifted telluric A
and B bands as they appear in the spectrum of G1 are also marked.

We see from its spectrum that G1 is probably best classified as a dwarf
elliptical, or possibly E/S0.  The two most prominent features in the
spectrum of G1 that suggest it is at redshift $z\approx0.22$ are the 4000
\AA\ break and the absorption feature that corresponds to the position
of the MgI$+$MgH band.  The narrow downward spike within this Mg band is
caused by residuals from subtracting the 6300 \AA\ night sky line. The
shape of the continuum between 4500 \AA\ and 8000 \AA\ matches the E-type
galaxy template spectrum remarkably well. However, given the spectrograph
slit size, which corresponds to $\approx$ 5 kpc in the blue and $\approx$
3 kpc in the red for $z=0.22$, a reasonable interpretation would also
be that the light being sampled by the spectrograph originates in the
{\it bulge} of an early type spiral galaxy. Template spectra of early
type spirals (S0 and S0/a) with non-Seyfert nuclei (Kennicutt 1992a)
also show the same characteristic continuum shape and spectral features
as the E galaxy spectrum shown on Figure 2. Emission lines, in particular
H$\alpha$, are weak in the integrated spectra of Sa galaxies and are not
present in earlier types, so the absence of H$\alpha$ and other emission
lines in the spectrum of G1 also strongly indicates that G1 is an early
type galaxy.  The expected position of H$\alpha$ at $z=0.22$ is marked
on Figure 2.  Kennicutt (1992b) points out that the equivalent width of
[OII]$\lambda$3727 emission in galaxies is typically one-third that of
H$\alpha$, so the absence of [OII] (which is in the most noisy part of
the spectrum) is consistent with the absence of H$\alpha$; moreover,
when H$\alpha$ is weak (as is the case here) the equivalent width of
[OIII]$\lambda$5007 rarely exceeds $\approx 1-2$ \AA, so it would also
not be expected.

It is interesting that some dwarf Es, E/S0s, and S0s are known to
have substantial \ion H1 halos (e.g. Lake, Schommer \& van Gorkom 1987
and Sadler et al. 2000), extending up to $\approx$ 5 times the optical
radius. The \ion H1 column densities in the inner contours easily exceed
the 2$\times10^{20}$ atoms cm$^{-2}$ requirement for producing a classical
\DLya absorption line.  Unlike the luminous Es and S0s, these early-type
dwarf galaxies are \ion H1-rich for their morphological class, with centrally
concentrated neutral gas and \ion H1 masses of a few times $10^8-10^9$
M$_{\odot}$.  However, these early-type \ion H1-rich dwarfs are evidently
rare locally.

The apparent magnitude and spectrum of G1 is reminiscent of some of
the faintest galaxies discovered in the Hawaii K-band survey of Songaila
et al.  (1994, e.g., see their object number 265 in field SSA22).  Indeed,
examination of the statistical results on the Hawaii deep survey fields
presented by Cowie et al. (1996) indicates that G1 is intrinsically one
of the faintest galaxies near redshift $z\approx0.22$ in comparison to
their sample.

We note that the spectrum of G1 recently obtained by Cohen (2000) is
consistent with our interpretation.

\medskip

\centerline{\it 3.4.1.2. The Spectral Energy Distribution}

\smallskip

Since our imaging data were taken with three different telescopes
in various conditions and with widely varying limiting magnitudes, an
appropriate procedure had to be used to accurately determine G1's colors.
The worst seeing conditions ($\approx 1.0$ arcsec) were in the U and
H bands.  Therefore, all of the other images (BRIJK) were degraded to
this resolution by convolution with a Gaussian before the photometric
measurements were made. An aperture size of 1.4 arcsec ($\approx 4.9$
kpc at $z=0.22$) was then determined to be a good choice to include most
of the galactic light detectable above the sky background in the faintest
image and still yield minimal errors. However, we note that nearly half of
G1's light in the WIYN R-band was not included in the 1.4 arcsec aperture
using this method. Extrapolation was then used to derive total magnitudes.
Comparison of the WIYN B-band and R-band images, which represents our
best two-color data, showed no {\it significant} evidence for a color
gradient in G1.  The estimated errors in G1's magnitudes (Table 2)
include both the statistical errors and zero-point calibration errors.

We see that G1 is a neutral-colored dwarf galaxy (B$-$K$ = 4.9$)
with $L_K = 0.10L^*_K$, $L_R = 0.12L^*_R$ and $L_B = 0.10L^*_B$, where an
$L^*$ galaxy corresponds to $M^*_K = -24.5$, $M^*_R = -21.9$ and $M^*_B =
-20.9$ (see Loveday 2000 for $M_K^*$, Marinoni et al. 2000 for $M_B^*$,
and Poggianti 1997 and Cowie et al. 1994 for K corrections).

Given the UBRIJHK colors of G1, we point out that further evidence of 
its redshift comes from a derivation of its photometric redshift. As
Figure 3 shows, $z=0.2$ is the most probable redshift based on
fitting 1996 Bruzual and Charlot galaxy model templates to the colors
(see more discussion below). In particular, it was possible to obtain
an excellent match to the spectral energy distribution of G1 using a
spectral synthesis code to model a galaxy near redshift $z=0.22$. While
we considered a variety of metal abundances, the fits were not very
sensitive to metallicity for these colors. However, slightly higher
and lower redshifts could not be ruled out definitively based on the
photometry alone (Figure 3) and, in the end, the best evidence for the
$z=0.22$ redshift of G1 comes from its slit spectrum (Figure 2) and
the circumstantial evidence provided by the existence of the $z=0.2212$
absorber in the field. Therefore, while the derivation of photometric
redshifts for intrinsically faint \DLya galaxies would seem to be a
promising method when used with the expectation that there will be a
galaxy in the field at the absorption redshift, the technique should
be used with caution since most photometric redshift techniques have
been tested by comparing photometric and spectroscopic redshifts of
intrinsically luminous galaxies, but evidence (\S4) suggests that \DLya
galaxies are often not very luminous.

In order to consider how well the observed colors of G1 at $z=0.22$
match different types of galaxies, and what this might tell us about the
present nature of G1 and its progenitor, we used the publicly available
photometric redshift code {\it hyperz} (Bolzonella, Pello \& Miralles
2000) to produce spectral synthesis models. First we dereddened the
colors of G1 using a Galactic extinction law and A$_B=0.18$ 
(Cardelli, Clayton, \& Mathis 1989).
We then used {\it hyperz} to investigate the range of models and ages
that produced acceptable fits to the observed colors of G1.  For our
analysis we set the metallicity to 20\% of the solar value. The amount of
intrinsic visual extinction, A$_V$, was allowed to be a free parameter,
assuming that the Calzetti et al. (2000) reddening law holds. The available
GISSEL96 template sets were for an instantaneous single-burst model,
plus models with exponentially decreasing star formation rates with
e-folding times in Gyrs of $\tau=$ 1, 2, 3, 5, 15, 30, and infinity
(i.e. constant star formation rate). Models with $\tau<3$ are generally
taken to correspond to early type galaxies (e.g. type E or S0).
Of these models, the burst models at ages between $0.36-3.5$
Gyrs, and with A$_V$ ranging between $1.3-0.1$, respectively,
were found to give acceptable fits to the color data. Also, a $\tau=1$
model with A$_V=0.5$ fits the color data reasonably well at
an age of $3.5$ Gyrs. A 0.72 Gyr old burst model with A$_V=1.1$
mag gave the best overall fit. [We note that for the Cardelli et al.
(1989) Galactic extinction law, A$_V=1$ corresponds to 
$N_{HI}=1.9\times10^{21}$ atoms cm$^{-2}$; while for the Calzetti et al. 
(2000) extinction law which we used in the {\it hyperz} simulations, 
A$_V=1$ might lie in the range $N_{HI}=0.4-2.0\times10^{21}$ atoms cm$^{-2}$,
but a good estimate is hindered by the absence of a metallicity measurement 
for G1. This should be compared to the value observed in the $z=0.22$ absorber
at impact parameter $b=20$ kpc, $N_{HI}=1.5\times10^{21}$ atoms 
cm$^{-2}$.] Figure 4 shows the color data with the
models over-plotted.  The best-fitting model is in panel (a) and three
reasonable alternative models are shown in panels (b), (c), and (d).
These galaxy templates are generally consistent with the observed
spectrum of G1 (Figure 2). A comparison of the model fits shown in Figure
4 indicates that far-UV observations could break some of the UBRIJHK
color degeneracy which makes it hard to distinguish age and extinction in
burst models.  Thus, lacking far-UV photometry of G1, we can
only infer that the progenitor population originated at a formation epoch
between $z_f\approx0.26$ (corresponding to the 0.36 Gyr old burst model)
and $z_f\approx0.84$ (corresponding to the 3.5 Gyr old models).  For the
four models illustrated in Figure 4, we show in Figure 5 the corresponding
inferred color (B$-$K) evolutionary history of G1's progenitor prior to its
current state at $z=0.22$, including effects due to the K-correction but
assuming that no change in intrinsic dust extinction occurred.  As would
be expected, in all cases the color evolution suggests that the progenitor
of G1, with B$-$K$=4.9$ at $z=0.22$, would have been considerably
bluer close to its formation epoch, i.e., B$-$K$<$4 at $z \approx z_f$.
Therefore, we conclude that G1 might be a faded, low-redshift counterpart
originating from the population of moderate-redshift ($z\approx0.3-0.9$)
``faint blue galaxies'' that have been identified in surveys (e.g. Ellis
1997 and references therein). Further details are given in the 
next section.

\medskip

\centerline{\it 3.4.1.3. The Star Formation Rate}

\smallskip

An upper limit to the H$\alpha$ flux in the spectrum of G1 (Figure 2)
has been estimated. This was done by first correcting the continuum 
shape near 8000 \AA\ to match that of the Figure 2 comparison spectrum. 
The absolute flux zero point was then obtained by adjusting
this corrected, standard-star-calibrated spectrum by a constant scaling
factor to make it agree with the more accurate optical photometric
measurements. Measurement of the spectrum using a slit corresponding
to a width of $\approx 3$ kpc near H$\alpha$ yields a 4$\sigma$ upper
limit on the rest frame H$\alpha$ flux of $f_{H\alpha} < 2.5\times 10^{-16}$
ergs cm$^{-2}$ s$^{-1}$, as measured over an $\approx$ 12 \AA\ interval.
At $z=0.22$ this translates to an H$\alpha$ luminosity of
$L_{H\alpha} < 3.4 \times 10^{40}$ ergs s$^{-1}$ and a corresponding star
formation rate $< 0.3$  M$_{\odot}$ yr$^{-1}$ (Kennicutt 1992b). This
limit is not unexpected for the bulge-dominated region of a galaxy.

We can also specify the details of the star formation process that
correspond to the spectral synthesis models which were discussed in the
previous section and illustrated in Figures 4 and 5. Using the total
magnitudes given in Table 2 we find that: (a) for the best-fitting 0.72
Gyr old burst model there is (of course) no current star formation and
the total stellar mass involved is $5.8\times10^9$ M$_{\odot}$, (b) for the
0.36 Gyr old burst model the total mass is $5.3\times10^9$ M$_{\odot}$,
(c) for the 3.5 Gyr old burst model the total mass is $7.1\times10^9$
M$_{\odot}$, and (d) for the 3.5 Gyr old $\tau=1$ model the total mass
is $7.6\times10^9$ M$_{\odot}$, where the initial star formation rate
is 7.8 M$_{\odot}$ yr$^{-1}$ and the star formation rate at $z=0.22$ 
is 0.24 M$_{\odot}$ yr$^{-1}$. This is consistent with our upper limit 
on the star formation of 0.3 M$_{\odot}$ yr$^{-1}$ from H$\alpha$. We
note that using total magnitudes (Table 1) results in stellar masses and 
star formation rates that are about a factor of two larger than when aperture 
magnitudes are used.

\medskip

\centerline{\it 3.4.1.4. The Radial Brightness Profile}

\smallskip

The radial brightness profile of G1 is also of interest. Using the WIYN
R-band data with the best seeing ($\approx 0.55$ arcsec), we analyzed
the radial brightness profile of G1.  The observed light profile is the
convolution of the true, two-dimensional light profile and the image
point-spread function.  If the true profile is symmetric and the PSF is
well approximated by a Gaussian, the convolution reduces to a simple
one-dimensional integral (see Binney \& Tremain 1987) that makes it
possible to compare a model radial brightness profile to the data in
a straightforward manner.  The observed light profile of G1 is very
close to symmetric, so we used this method.  Elliptical isophotes were
fitted to the brightness profile of G1. The resulting radial brightness
profile along the semi-major axis is shown in Figure 6. Three models were
fitted to the data: (a) a pure r$^{1/4}$ (bulge) profile, (b) a pure
exponential (disk) profile, and (c) a combination bulge-disk profile.
As shown in the corresponding panels (a-c) of Figure 6, a pure-disk
model is a poor fit, while a pure-bulge model gives better results. The
bulge-disk combination is clearly the best fit and is the preferred model,
but it includes an extra free parameter. This simple model and the small
angular extent of G1 compared to the seeing make it difficult to pin
down the bulge-to-total light ratio, the bulge half-light radius or the
disk scale-length with accuracy, but our fits suggest that a moderate
amount of each is necessary to fit the light profile well.

Owing to the need for some disk light, this result therefore seems
reasonably consistent with the interpretation that G1 is an early-type
``dwarf spiral,'' which has been discussed as a recently-recognized new
class of galaxy (Schombert et al. 1995).  In many respects these galaxies
resemble the \ion H1-rich dwarf Es, E/S0s, and S0s mentioned in \S3.4.1.1.
According to Schombert et al., the dwarf spirals are found only in the
field and have total luminosities $M_B>-17.6$, diameters $R_{26}<6.5$
kpc, low central surface brightnesses $\mu_0 > 24$ B mag arcsec$^{-2}$,
and low \ion H1 masses $M_{HI}\leq 1.7 \times 10^9$ M$_{\odot}$ (converting the
Schombert et al. values to our assumed cosmology with $H_0=65$ km s$^{-1}$
Mpc$^{-1}$). This represents a class of objects which are currently rich
in neutral gas, small in angular size, and low in surface brightness. The
central surface brightness of G1 ($\mu_0 \approx 23.4$ B mag arcsec$^{-2}$)
is somewhat brighter than the parameters given by Schombert et al.

\medskip

\centerline{\it 3.4.1.5. The \ion H1 Mass Estimate}

\smallskip

G1 is displaced 5.7 arcsec from the quasar sight-line, corresponding to
an impact parameter of $b=20$ kpc.  This is a relatively large impact
parameter in relation to the apparent optical size of G1 (Figure 1).
It is not unusual to find evidence for a relatively large impact parameter
when trying to identify galaxies giving rise to quasar absorption-line
systems, but it is somewhat surprising for a \DLya system and so it leaves
open the possibility that the site of the absorption is an \ion H1 cloud in
a galaxy that is a companion to G1. However, consideration of this is
beyond the scope of the available data, so we have only concentrated on
the description of G1 here.  We note that if the $z=0.2212$ absorber
is assumed to have a uniform \ion H1 column density of $\approx 7 \times
10^{20}$ atoms cm$^{-2}$ and extend over a radius equal to
the impact parameter, then its deduced \ion H1 mass is $M_{HI}$ $\approx 2.8
\times 10^{10}$ M$_{\odot}$. This \ion H1 mass estimate is about an order
of magnitude greater than quoted above for \ion H1-rich dwarf Es, E/S0s,
S0s, or spirals.  

\subsubsection{The ``Fuzz'' Near the Quasar ($z=0.09$)}

The arm and jet are the two brightest patches of low surface brightness
extended light near the quasar (i.e. the ``fuzz''). Since we do not
have a useful spectrum of any part of the fuzz, or data that can be used
to derive a photometric redshift for these features, their
redshifts cannot be known with certainty. In principal, the light could
be from a combination of: (1) unrelated foreground or background objects
at neither of the two damped redshifts, (2) the quasar host galaxy at
$z=0.63$, (3) the ``true'' \DLya galaxy at $z=0.22$ (assuming that G1
at redshift $z=0.22$ does not have a sufficient \ion H1 extent), and
(4) the \DLya galaxy at $z=0.09$. Examination of the infrared images do
suggest a combination of objects or perhaps a single disturbed object (see
Figure 1). Therefore, we will discuss each of these four possibilities.

(1) Contamination from unrelated objects is possible. However,
examination of the infrared images suggests that $<1$\% of
the frame contains detectable infrared sources at the sensitivity of
the K-band images. Therefore, if an unrelated object or objects are
present near the quasar sight-line, this is an unlucky configuration,
so we will not consider this possibility further.

(2) There is a reasonable chance that some of the quasar's host galaxy
light  at $z=0.63$ is present in the infrared light surrounding the
quasar. The quasar, OI$\;$363, is optically luminous (V$\approx$16.1,
M$_V\approx-26.1)$ and radio loud. Infrared studies of optically luminous,
radio loud quasars in the H-band (McLeod \& Reike 1994) show that it is
not unusual for these types of quasars to have luminous host galaxies,
e.g., $> 2L_H^*$, and HST-WFPC2 studies at $z<0.46$ show that the
host galaxies of radio loud quasars are almost exclusively ellipticals
(Hamilton, Casertano, \& Turnshek 2000). Examination of the infrared
OI$\;$363 images after subtraction of the unresolved nuclear quasar light
(Figure 1) indicates that the remaining light is not exclusively from an
elliptical host centered on the quasar.  A reasonable deconvolution of the
image suggests that the nuclear quasar component has $M_K \approx -28.6$
($L_K \approx 44L^*_K$) while its host has an $r^{1/4}$ profile and $M_K
\approx -25.8$ ($L_K \approx 3.3L^*_K$), which is completely consistent
with previous findings on radio loud quasar hosts. Nevertheless, after
subtraction of the nuclear and host components some light remains between
the arm and the jet. In any case, the mere presence of the arm and the
jet indicates that the surrounding light is not exclusively associated
with an elliptical host.  An illustration of the light that we believe
is dominated by the quasar elliptical host is contained within the
ellipse drawn on Figure 1a.  However, there are uncertainties with this
deconvolution and better spatial resolution would be needed to accurately
separate out the various components.

(3) The possibility that some of the extended, low surface brightness
light surrounding the quasar is due to the \DLya galaxy at $z=0.22$
is dismissed since G1 is identified as the \DLya galaxy.  However,
recall that the impact parameter of G1 is 5.7 arcsec or $b=20$ 
kpc, which may be large for a dwarf galaxy giving rise to \DLya 
absorption (\S3.4.1.5).

(4) From the arguments presented above (1-3), we conclude that {\it some}
of the light surrounding the quasar is from the \DLya galaxy at $z=0.09$.

Owing to the approximately circular nature of the resolved light,
regardless of whether a contribution from an elliptical host is removed,
and the possibility of an actual spiral arm, it may be that this \DLya
galaxy is a nearly face-on spiral. If this turns out to be the case,
the low surface brightness dwarf spiral interpretation might in fact be
more descriptive of the $z=0.09$ \DLya galaxy than for G1 at $z=0.22$
(\S3.4.1.4). At the same time, the considerable morphological structure
surrounding the quasar leaves open the possibility that the $z=0.09$
\DLya galaxy is an irregular or an interacting system.  In any case,
if we want to place a conservative {\it upper limit} on the light from
the $z=0.09$ \DLya galaxy it is appropriate to assume that {\it all} of
the light surrounding the unresolved part of the quasar image (including
what we believe to be the quasar host) belongs to it. With this assumption
we obtain an upper limit on the K-band luminosity of the \DLya galaxy at
$z=0.09$ of $L_K \approx 0.13L^*_K$.  Assuming the arm and jet are both
at $z=0.09$, the arm has $L_K \approx 0.0007L^*_K$ with color R$-$K$>$5.5
(i.e. it is red and not detected in our R-band WIYN data) and the jet
has $L_K \approx 0.0005L^*_K$ with a blue color of R$-$K$\approx$3.2.
This would imply that the jet is a site of relatively recent star
formation, while the arm is an older population. We estimate the diameter
of the measurable light distribution to be $\approx 7-9$ arcsec, or
$\approx 12-16$ kpc, and the separation from the quasar sight-line to
be $<2$ arcsec, or impact parameter $b<3.6$ kpc.  As noted above, light
contributed by an elliptical quasar host galaxy (i.e. some light inside
the ellipse shown on Figure 1a) might have to be subtracted from the
measured upper limit on the \DLya galaxy's light in order to determine
its true luminosity. In fact, a reasonable deconvolution suggests that
the quasar host contributes $\approx 50$\% of the total measured light,
in which case the luminosity of the $z=0.09$ \DLya galaxy is $L_K \approx
0.08L^*_K$.  Regardless of the amount of contribution from the host
galaxy, the resulting interpretation is that the $z=0.09$ \DLya galaxy is
a low surface brightness galaxy with a diameter of $\approx 14$ kpc at low
impact parameter from the quasar sight-line. The details cannot be sorted
out definitively without better data.  Thus, the morphology of this low
surface brightness dwarf galaxy remains uncertain $-$ (another?) dwarf
(possibly face-on) spiral, or an irregular, or an interacting system.

Given the above indication of the putative $z=0.09$ \DLya galaxy's
diameter, it is useful to estimate its \ion H1 gas mass using some
reasonable assumptions. If we simply assume that the circular \ion H1 gas
radius is the same as the extent of the galaxy seen in the K-band image
(i.e. radius r $\approx$ 7 kpc) and that the \ion H1 column density is
uniformly $N_{HI} \approx 1.5 \times 10^{21}$ atoms cm$^{-2}$ over that
extent, the deduced \ion H1 mass is $M_{HI} \approx 6.5 \times 10^9$
M$_{\odot}$.  This is a factor of 2 larger than the upper limit of Lane
et al. (2000b) but, as discussed in \S2.1, some of the \ion H1 21 cm
emission searched for by Lane et al. (2000b) may have been missed since
velocity intervals which included absorption had to be excluded in their
analysis. At the same time, the assumptions used above may over-estimate
the \ion H1 mass since the impact parameter is evidently small and the
\ion H1 gas distribution may be patchy or fall off exponentially with
impact parameter.  Nevertheless, the order of magnitude of this estimate
is of interest; the total \ion H1 mass is smaller than our estimate for
the $z=0.2212$ \DLya system (\S3.4.1.5).

\section{DISCUSSION}

Although there are still open questions which better data could address,
the findings discussed here represent some of the first results on
the {\it detailed properties} of \DLya galaxies. Perhaps fittingly,
these results were determined from the two lowest redshift ($z=0.09$
and $0.22$) confirmed classical \DLya systems known. 

The results of our analysis indicate the following. The $z=0.22$ \DLya
galaxy is a faint (B$=22.7$, K$=17.8$) neutral-colored (B$-$K$=4.9$)
dwarf galaxy with K- and B-band luminosities of $L_K = 0.10L_K^*$
and $L_B = 0.10L_B^*$, respectively.  Its spectrum is that of an
early type galaxy (S0 or even E).  Analysis of the 7-color (UBRIJHK)
data suggests that G1's spectral energy distribution is well-fitted
by an instantaneous single-burst model with the burst taking place
0.36 (A$_V = 1.3$) to  3.5 (A$_V = 0.1$) Gyrs ago, or a 3.5 Gyr old
model with an exponentially decreasing star formation rate ($\tau=1$
Gyr) and A$_V =0.5$.  Its current rate of star formation is low, $<
0.3$ M$_{\odot}$ yr$^{-1}$. Examination of the evolutionary history of
these models leads us to conclude that the progenitor population of G1,
if seen near the population's formation epoch at $z_f\approx0.3-0.9$,
would have properties similar to the population of ``faint blue galaxies''
presently observed at moderate redshift.  Analysis of the extended radial
light profile from the WIYN R-band image of G1 indicates that its outer
isophotes are better fit by an exponential than an r$^{1/4}$ law, which
offers some evidence for an early-type dwarf S0 or spiral interpretation.
The putative $z=0.09$ \DLya galaxy is likely to be all or part of the
resolved light surrounding the quasar. It is also a dwarf galaxy, and is
inferred to have $L_K \le (0.08-0.13)L^*_K$, with the higher value holding
for the case where the quasar host galaxy at $z=0.63$ does not contribute
to the detected resolved light.  There is considerable morphological
structure in the light surrounding the quasar, including what we have
called an arm and a jet. The arm is red in color and if it is an {\it
actual} arm, this suggests that the \DLya galaxy might be a low surface
brightness dwarf spiral, but the presence of the other structure (i.e. a
blue colored jet with apparent structure) leaves open the possibility
that the \DLya galaxy is an irregular or an interacting system. In any
case, the $z=0.22$ \DLya galaxy is more compact, while the inferred \DLya
galaxy at $z=0.09$ has low surface brightness. For these galaxies the
\ion H1 gas masses deduced from the observed impact parameters and \ion
H1 column densities lie in the range $M_{HI} \approx (6-23)\times10^{9}$
M$_{\odot}$. These estimates are crude because we do not know if the
observed \ion H1 is distributed like a face-on circular slab around the
center of the identified \DLya galaxy (like our calculation of the \ion
H1 mass assumed) or whether the \ion H1 gas is exponentially distributed
or clumped in giant clouds associated with the identified galaxy.

In any case, it would clearly have been much more difficult to deduce
these properties from similar observations if these same objects were
at moderate to high redshift.  However, a collection of such detailed
results, used in conjunction with statistical results on the incidence,
cosmological \ion H1 gas mass density and column density distribution of
\DLya absorption systems (Rao \& Turnshek 2000), will eventually be
needed to accurately reveal the evolution of the bulk of the neutral
gas in the Universe (as tracked by \DLya quasar absorption systems)
and the corresponding evolutionary relationship to \DLya galaxies
(i.e. the associated stellar light).  One qualitative question presently
of interest is to determine if the \DLya population is slowly evolving
from $z\approx4.0$ to $z\approx0.5$, as has been recently suggested by
Pettini et al. (1999) and Rao \& Turnshek (2000) on the basis of their
metallicities and statistics, respectively.  It would seem that there has
to be some significant evolution in the \DLya population from high to low
redshift if the results of Rao \& Briggs (1993) at $z=0$ are correct. The
Rao \& Briggs (1993) study places almost all of the local \ion H1 gas in
luminous spirals. 
Present evidence suggests that luminous galaxies were approximately in
place by $z\approx1$, with less luminous galaxies showing more evolution
at lower redshift (e.g. Im et al. 1996, 1999).  Therefore, one could envisage a scenario
where luminous galaxies contribute little to the \DLya detections,
except locally, while an evolving population of dwarfs and low surface
brightness galaxies cause the bulk of the detections at $z>0$. Indeed,
the present searches for low-redshift DLA galaxies (Le Brun et al 1997;
Turnshek et al. 2000) suggest that a significant fraction of them are
dwarf and low surface brightness galaxies, and the possible connection
to the evolving moderate redshift ``faint blue galaxy'' population
mentioned earlier is intriguing.  Clearly the statistical survey
results and the low-redshift follow-up studies of DLA systems need to
be extended and improved to address these evolutionary issues properly.
It is also of interest to consider selection effects more
carefully as well as design experiments which will reach to 
lower \ion H1 masses in searches for large quantities of \ion H1 gas not associated
with luminous spirals. The current 21 cm emission surveys are sensitive
to $M_{HI} \approx 10^7$ M$_{\odot}$ locally. 

Regardless of the outcome of such new programs, the results reported
here are consistent with our earlier findings (RT98) and serve to
emphasize that \DLya absorbing galaxies are not exclusively 
luminous spirals.  They are drawn from a wide variety of types of
gas-rich galaxies.

\medskip

\centerline{\bf Acknowledgements}

\smallskip

We wish to thank Dr. Frank Briggs for collaboration on parts of this
program and Drs. Bolzonella, Pello, and Miralles for making their spectral
synthesis code {\it hyperz} publicly available. We wish to also thank
the support astronomers at the IRTF, Mayall, MDM, and WIYN telescopes
for their assistance. This work was supported in part by an NSF grant
and a NASA-LTSA grant.

\newpage

\begin{figure}
\centerline{\epsfig{file=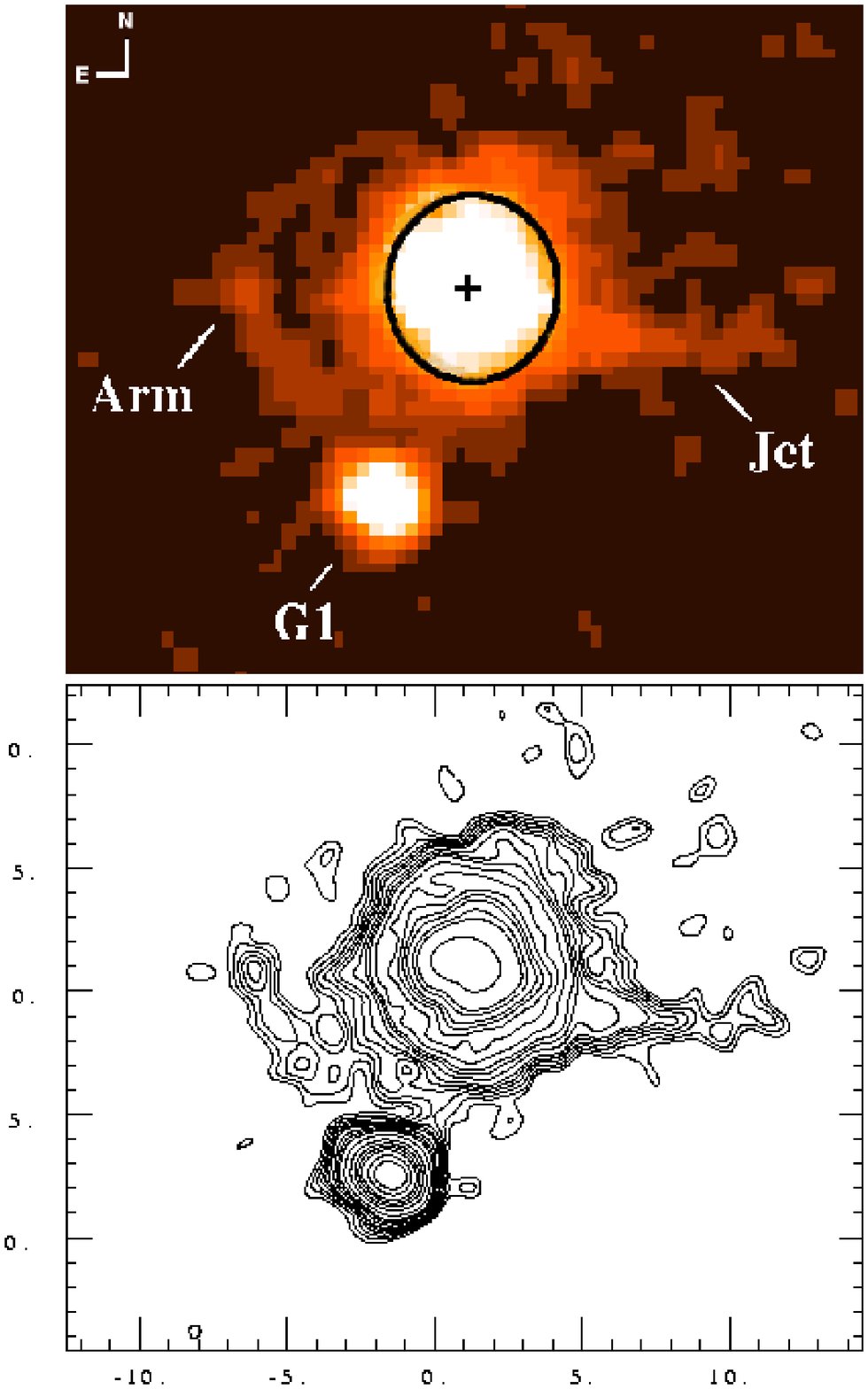,height=6.0in}}
\caption{The top panel shows an IRTF K-band image of the
OI$\;$363 quasar field after smoothing with a Gaussian having a FWHM
equivalent to the seeing (0.75 arcsec) and then subtracting the
quasar's nuclear PSF. The center of the subtracted nuclear PSF is
marked with a ``+.''  The image is $20\times 17$ arcsec. 
In the central region, enclosed
in the ellipse drawn on the figure, the light appears to be dominated
by an elliptical host ($M_K \approx -25.8$ or $L_K \approx 3.3L_K^*$
at $z_{Q,host}=0.63$) to the quasar ($M_K \approx -28.6$
or $L_K \approx 44L_K^*$ at $z_{Q,nuc}=0.63$).  One or more of the low
surface brightness features labeled ``arm'' and ``jet'', and the light
between them (see text) is presumably due to the DLA galaxy at $z=0.09$,
corresponding to the DLA absorption-line system at $z=0.0912$ with
$N_{HI}=1.5\times10^{21}$ atoms cm$^{-2}$.  The dwarf galaxy labeled
G1 is at $z=0.22$, corresponding to the DLA absorption-line system at
$z=0.2212$ with $N_{HI}=7.9\times10^{20}$ atoms cm$^{-2}$. The bottom panel 
shows an isophotal plot of the same region where the faintest isophote
corresponds to 3$\sigma$ above the background.
}
\end{figure}

\newpage

\begin{figure}
\epsfig{file=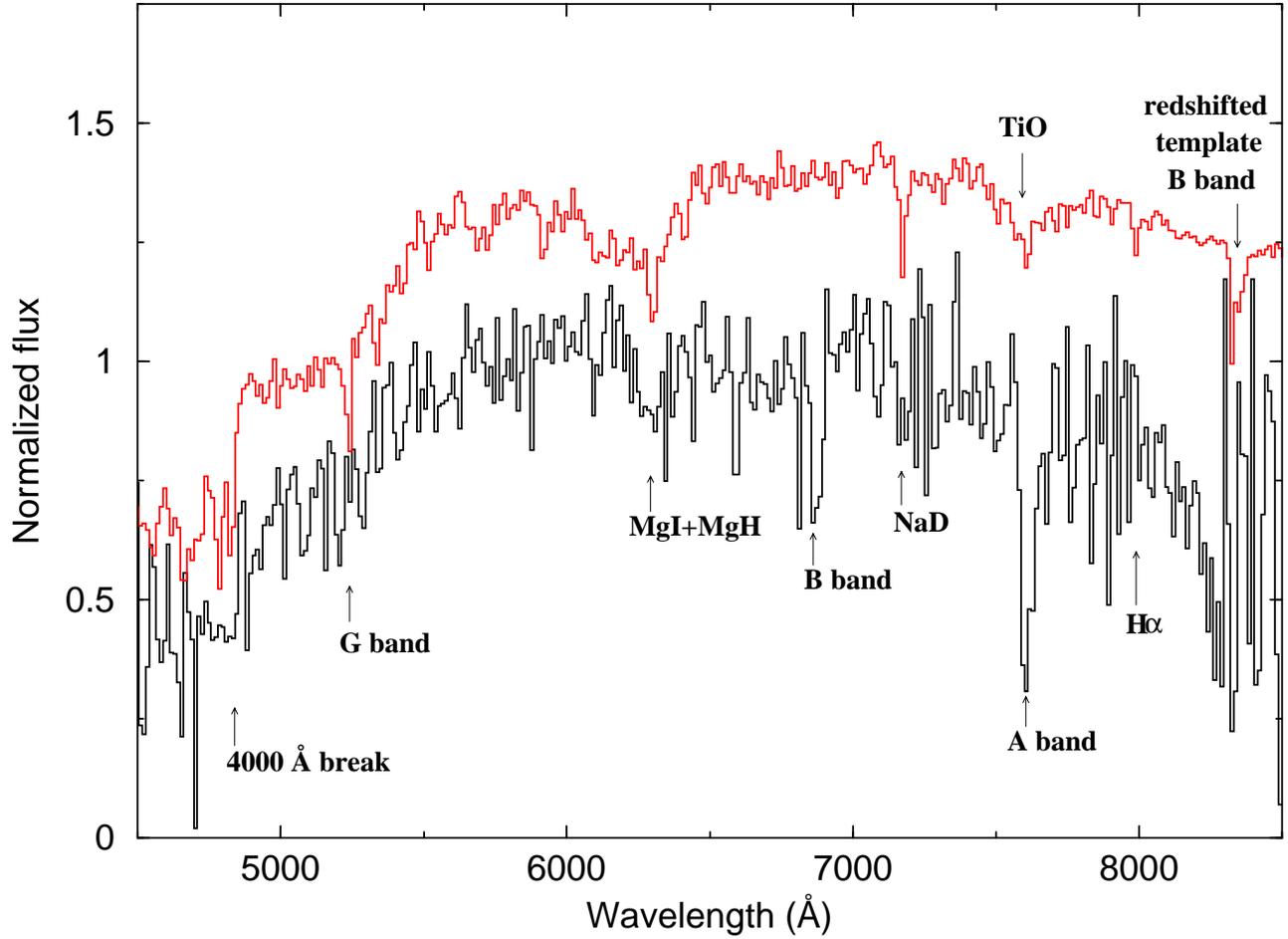,height=5.0in}
\caption{NOAO KPNO 4-m CRYOCAM spectrum of G1 with matching
early-type galaxy template spectrum at $z=0.22$ over-plotted. See
\S3.4.1.1 for a discussion of the spectrum.
}
\end{figure}

\newpage

\begin{figure}
\epsfig{file=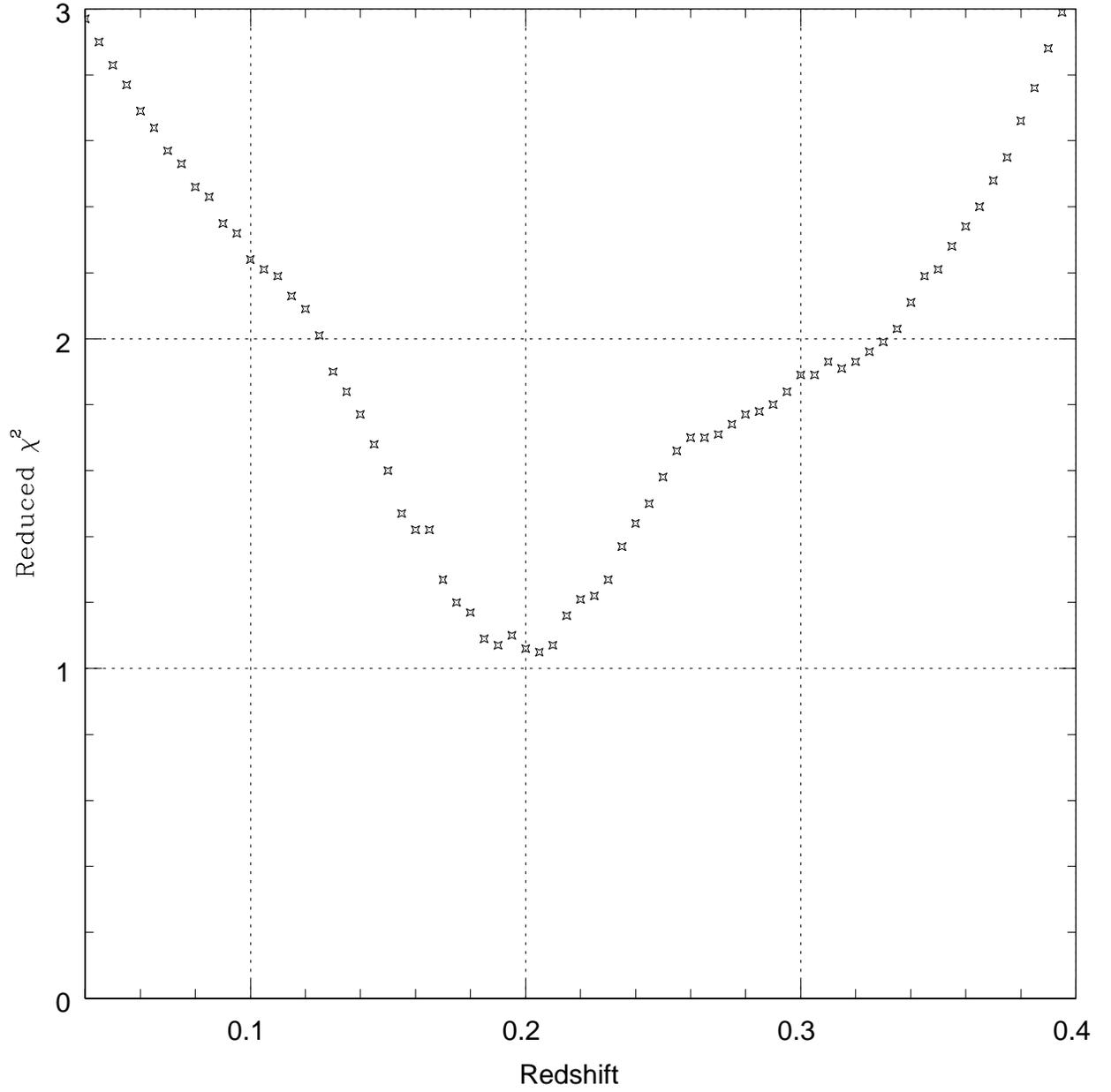,height=7.0in}
\caption{$\chi^2$ plot showing the reliability of the photometric
redshift determination for G1. The most probable photometric redshift
is $z\approx 0.2$, which matches the slit redshift.
}
\end{figure}

\newpage

\begin{figure}
\epsfig{file=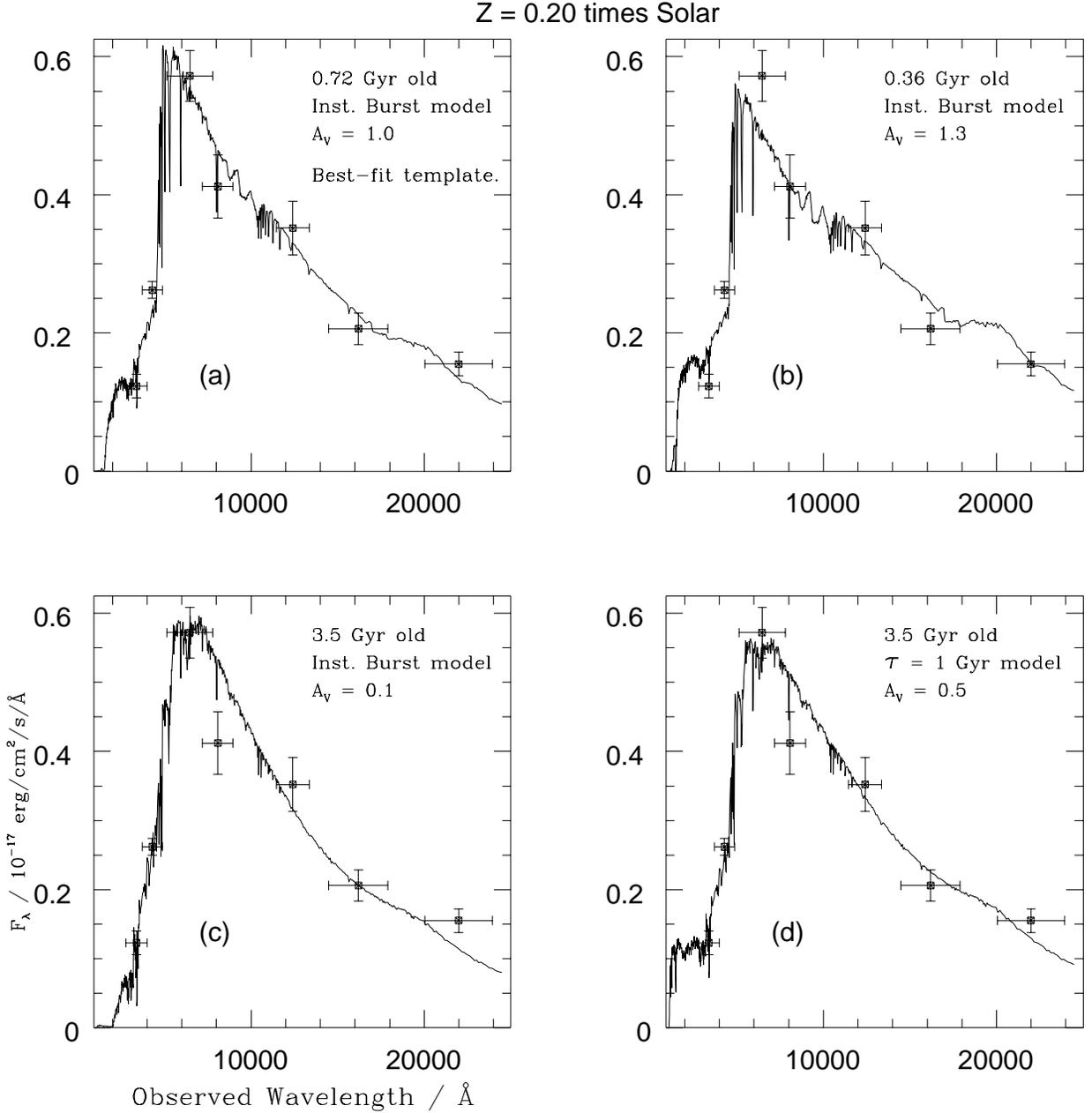,height=7.0in}
\caption{Photometry of G1 ($z=0.22$) corrected for Galactic
reddening (A$_B=0.18$) and converted to an f$_{\lambda}$ scale showing
the agreement between the colors (U, B, R, I, J, H, K) and the 1996
models of Bruzual and Charlot (see Bruzual \& Charlot 1993).  The four
panels are as follows:
(a) a 0.72 Gyr old instantaneous-burst model with A$_V=1.0$, 
(b) a 0.36 Gyr old instantaneous-burst model with A$_V=1.3$,
(c) a 3.5 Gyr old instantaneous-burst model with A$_V=0.1$, and 
(d) a 3.5 Gyr old model with an exponentially decreasing star 
formation rate ($\tau=1$ Gyr and A$_V=0.5$). All models are 20\%
solar metallicity and use a Scalo (1986) initial mass function.
Panel (a) is the best-fitting model, while the others are acceptable
alternatives.
}
\end{figure}

\newpage

\begin{figure}
\epsfig{file=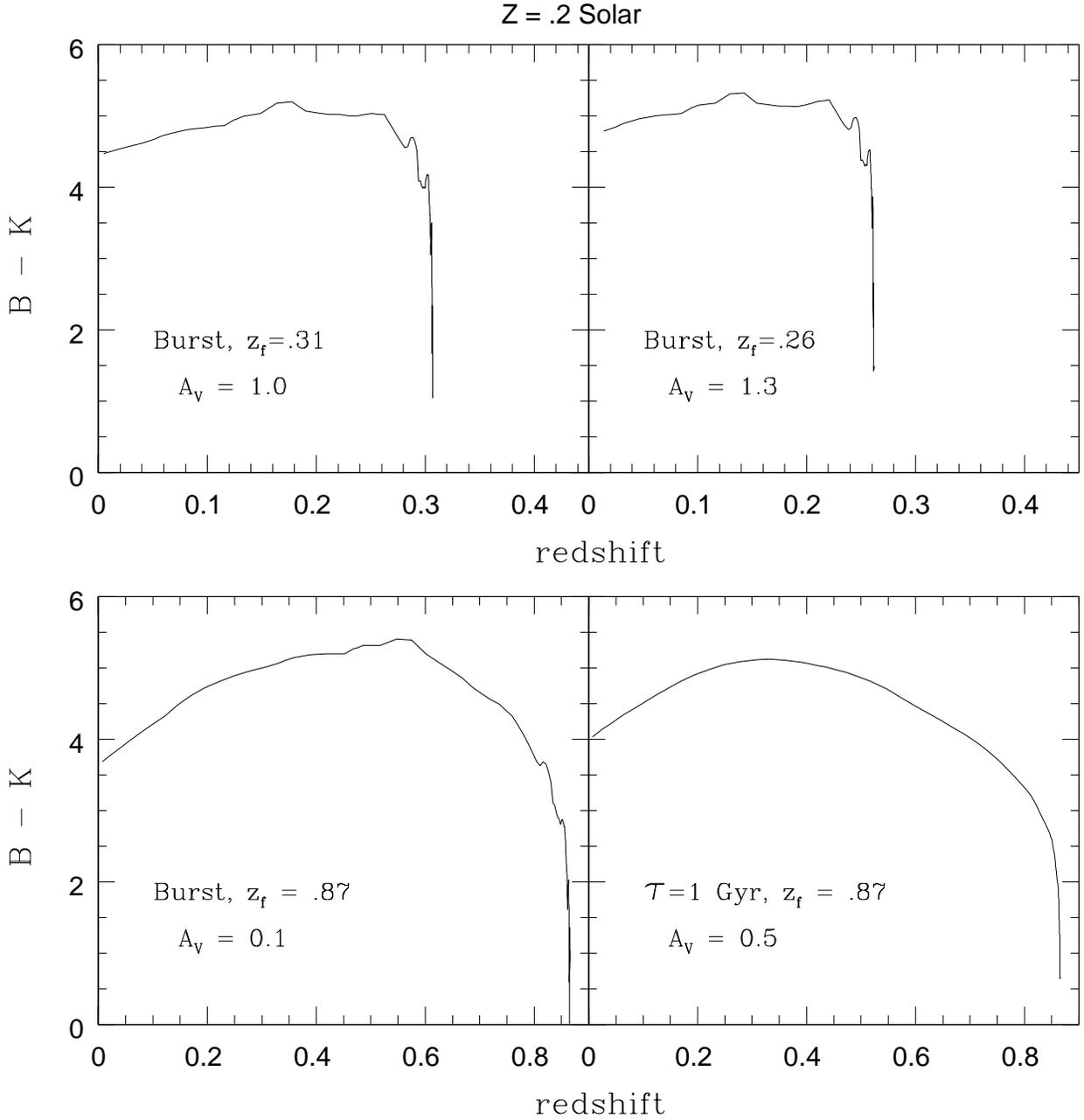,height=7.0in}
\caption{Calculated B$-$K color evolution of G1 ($z=0.22$) from 
its formation epoch to the present. Panels a-d correspond to panels a-d 
in Figure 4, respectively. Dust extinction was assumed constant in the 
calculation.
}
\end{figure}

\newpage

\begin{figure}
\epsfig{file=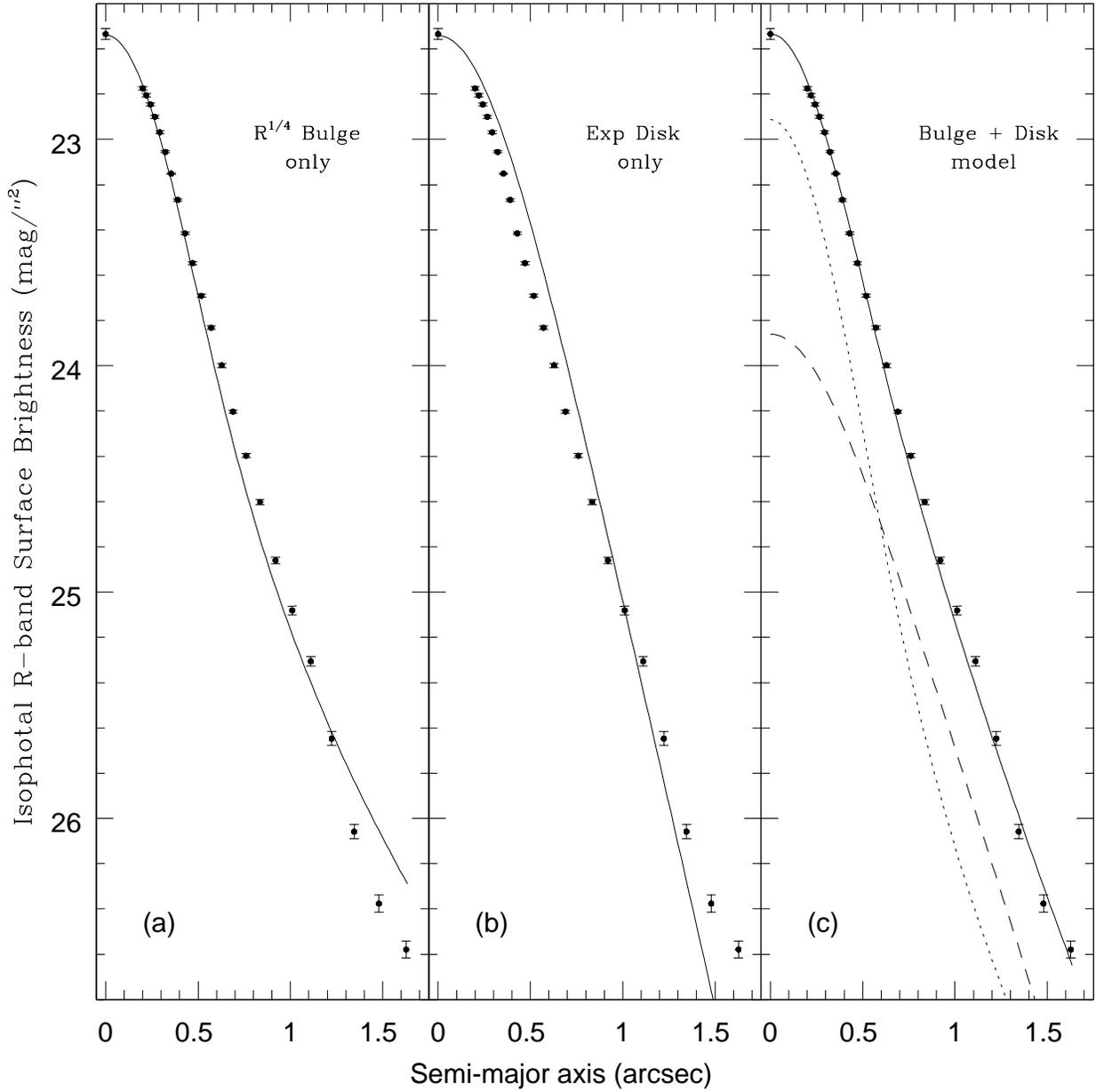,height=7.0in}
\caption{Model fits to the the R-band radial light profile of G1:
(a) an r$^{1/4}$ (bulge) profile alone, 
(b) an exponential (disk) profile alone, and 
(c) the best-fitting model where the
interior isophotes are dominated by an r$^{1/4}$ law and the outer isophotes
are dominated by an exponential.
}
\end{figure}

\newpage

\begin{deluxetable}{cclc}
\small
\tablewidth 4.0in
\tablecaption{Journal of Imaging Observations}
\tablehead{          \colhead{Filter}
                    & \colhead{Telescope}
                    & \colhead{Observation}
                    & \colhead{Exp. Time} \\ [.1ex]
                     \colhead{}
                    & \colhead{}
                    & \colhead{Date}
                    & \colhead{(m)}
                    }
\startdata
U   & MDM  & 1999 Feb 20 & 60 \\
 "  &  "   & 1999 Nov 13 & 90 \\
 "  &  "   & 1999 Nov 14 & 60 \\
B   & WIYN & 1997 Nov  5 & 30 \\
 "  &  "   & 1998 Nov 25 & 15 \\
R   &  "   & 1997 Nov  5 & 45 \\
 "  &  "   & 1999 Jan 18 & 15 \\
I   &  "   & 1997 Dec  5 & 50 \\
J   & IRTF & 1998 Dec 13 & 30 \\ 
 "  &  "   & 1999 Apr 29 & 63 \\
H   &  "   & 1998 Dec 13 & 30 \\
 "  &  "   & 2000 Mar  8 & 63 \\
K   &  "   & 1998 Dec 12 & 130\\
 "  &  "   & 1998 Dec 14 & 132\\
 "  &  "   & 1999 Apr 28 & 63 \\
 "  &  "   & 2000 Mar  6 & 126\nl
\enddata
\end{deluxetable}

\begin{center}
\begin{deluxetable}{ccccccc}
\small
\tablewidth 6.0in
\tablecaption{UBRIJHK Photometry\tablenotemark{a}}
\tablehead{           \colhead{}
                    & \colhead{Limiting}
                    & \colhead{G1}
                    & \colhead{G1}
                    & \colhead{G1}
                    & \colhead{Jet}
                    & \colhead{Arm} \\ [.1ex]
                    \colhead{Filter}
                    & \colhead{mag\tablenotemark{b}}
                    & \colhead{Aperture mag}
                    & \colhead{Total mag}
                    & \colhead{$\mu_0$\tablenotemark{c}}
                    & \colhead{$\mu$}
                    & \colhead{$\mu$}
                    }
\startdata
U & 24.7 & 23.67 (0.15)& 22.8 (0.15)& 23.8                  (0.3)& \nodata
& \nodata    \\
B & 25.7 & 23.46 (0.05)& 22.7 (0.05)& 23.4                  (0.2)& \nodata
& \nodata    \\
R & 27.0 & 21.49 (0.07)& 20.8 (0.05)& 21.5\tablenotemark{d} (0.1)& 25.1
(0.2)& \nodata    \\
I & 24.1 & 21.07 (0.12)& 20.4 (0.1) & 21.0                  (0.1)& \nodata
& \nodata    \\
J & 22.8 & 19.82 (0.12)& 18.9 (0.1) & 19.9                  (0.1)& 22.1
(0.5)& \nodata    \\
H & 21.0 & 19.38 (0.12)& 18.7 (0.1) & 19.6                  (0.1)& \nodata
& \nodata    \\
K & 22.5 & 18.52 (0.12)& 17.8 (0.1) & 18.3\tablenotemark{e} (0.1)& 21.9
(0.2)& 21.5 (0.2) \nl
\enddata
\tablenotetext{a}{corrected for Galactic extinction}
\tablenotetext{b}{2-$\sigma$ detection limit in 1.4 arcsec aperture}
\tablenotetext{c}{central surface brightness at 1 arcsec resolution}
\tablenotetext{d}{20.6 at 0.55 arcsec resolution}
\tablenotetext{e}{17.8 at 0.75 arcsec resolution}
\end{deluxetable}
\end{center}

\end{document}